\documentclass[pdflatex,sn-mathphys-num]{sn-jnl}

\usepackage{graphicx}%
\usepackage{multirow}%
\usepackage{amsmath,amssymb,amsfonts}%
\usepackage{amsthm}%
\usepackage{mathrsfs}%
\usepackage[title]{appendix}%
\usepackage{xcolor}%
\usepackage{textcomp}%
\usepackage{manyfoot}%
\usepackage{booktabs}%
\usepackage{algorithm}%
\usepackage{algorithmicx}%
\usepackage{algpseudocode}%
\usepackage{listings}%
\usepackage[normalem]{ulem}%
\usepackage{siunitx}
\usepackage[normalem]{ulem}

\unnumbered

\begin{document}

\title[Article Title]{Single Photon Emitters in Ultra-Thin Hexagonal Boron Nitride Layers}

\author[1]{\fnm{Le} \sur{Liu}}
\equalcont{These authors contributed equally to this work.}

\author[1]{\fnm{Igor} \sur{Khanonkin}}
\equalcont{These authors contributed equally to this work.}

\author[1]{\fnm{Johannes} \sur{Eberle}}
\equalcont{These authors contributed equally to this work.}

\author[1]{\fnm{Bernhard} \sur{Rizek}}

\author[1]{\fnm{Stefan} \sur{F\"{a}lt}}

\author[2]{\fnm{Kenji} \sur{Watanabe}}

\author[3]{\fnm{Takashi} \sur{Taniguchi}}

\author[1]{\fnm{Ata\c{c}}\sur{Imamo\u{g}lu}}

\author*[1]{\fnm{Martin} \sur{Kroner}}\email{mkroner@phys.ethz.ch}

\affil[1]{\orgdiv{Institute for Quantum Electronics}, \orgname{ETH Zurich}, \city{Zurich}, \country{Switzerland}}

\affil[2]{\orgdiv{Research Center for Electronic and Optical Materials, National Institute for Materials Science}, \orgname{NIMS}, \city{Tsukuba}, \country{Japan}}

\affil[3]{\orgdiv{Research Center for Materials Nanoarchitectonics, National Institute for Materials Science}, \orgname{NIMS}, \city{Tsukuba}, \country{Japan}}

\abstract{

Single-photon emitters (SPE) in hexagonal boron nitride (h-BN) are promising for applications ranging from single-photon sources to quantum sensors. Previous studies exclusively focused on the generation and characterization of SPEs in relatively thick h-BN layers {\unboldmath($\geq$ 30 nm)}. However, for electrical and magnetic sensing applications, the thickness of the h-BN limits the attainable spatial resolution. Here, we report the observation of blue-wavelength emitters (B-centers) activated by electron beam irradiation in ultra-thin {\unboldmath($\simeq$ 3 nm)} h-BN. These SPEs in ultra-thin flakes exhibit reduced brightness, broader zero-phonon line, and enhanced photobleaching. Remarkably, upon encapsulation in thicker h-BN, we restore their brightness, narrow linewidth (\qty{230}{\micro\eV} at \qty{5}{\kelvin}, resolution limited), suppress photobleaching, and confirm single-photon emission with {\unboldmath \( g^{(2)}(0) <  0.4\)} at room temperature. The possibility of generating SPEs in a few-layer h-BN and their subsequent incorporation into a van der Waals heterostructure paves the way for achieving quantum sensing with unprecedented nanometer-scale spatial resolution.}


\maketitle

\section{Introduction}\label{Introduction}

Single photon emitters (SPEs) in atomically thin van der Waals materials, such as transition-metal dichalcogenides (TMDs) \cite{srivastava2015optically,koperski2015single,he2015single,chakraborty2015voltage,klein2021engineering} or hexagonal boron nitride (h-BN) \cite{tran2016quantum,mendelson2021identifying,bourrellier2016bright,fournier2021position,gale2022site,gottscholl2020initialization,Mai2025review}, have emerged as a promising platform for sensing local electric \cite{akbari2022lifetime,zhigulin2023stark} and magnetic \cite{gottscholl2021spin,huang2022wide} fields. Moreover, they could be used to detect the compressibility of proximal correlated electrons~\cite{li2021imaging} due to the highly sensitive and tunable two-dimensional (2D) interface. The 2D nature allows for integration into heterostructures in close proximity to targeted quantum materials \cite{huang2022wide,zhou2024sensing,Ren2024,Durand2023}. Combining the atomic nature of quantum emitters, this platform provides a possibility to image quantum phases \cite{smolenski2021signatures,li2021imaging} emerging in 2D systems at the sub-nanometer scale.

Among the numerous defects in van der Waals materials, blue centers (B-centers) in h-BN have gained significant attention due to their exceptional brightness \cite{fournier2021position,gale2022site}, room temperature optical stability \cite{fournier2021position,gale2022site,chen2023annealing}, narrow linewidth of the zero phonon line (ZPL) around \qty{436}{\nano\meter} \cite{fournier2023investigating,zhigulin2023stark},
and large polarizability by external electric fields \cite{zhigulin2023stark}. Recently, the indistinguishability of photons emitted by a B-center has been demonstrated \cite{gerard2025}. Another advantage of these defects is that they can be created simply by electron irradiation, e.g, using a scanning electron microscope (SEM) \cite{fournier2021position,gale2022site,roux2022cathodoluminescence}. However, this technique has so far only been demonstrated to work for the generation of B-centers in h-BN flakes of at least \qty{30}{\nm} thickness. 
This poses a severe limitation on the applicability of B-centers for sensing applications, where the defect needs to be situated as close as possible to the van der Waals surface of the h-BN crystal, to ensure the highest possible sensitivity and spatial resolution \cite{Ren2024, Kubanek2022}. The possibility of creating B-centers in pristine, ultra-thin h-BN (thickness $<$ \qty{5}{\nano\meter}) has been explored \cite{nedic2024electron}, albeit unsuccessfully.

In this Letter, we address this challenge by demonstrating the successful creation of B-centers in ultra-thin h-BN with a thickness as low as \qty{3}{\nm}. We observe that the standard electron beam irradiation technique is not suitable for thin h-BN flakes, as the surface damage caused by electron beam irradiation becomes significant when the thickness reaches the atomic layer limit. Moreover, hydrocarbon contamination forms on the surface areas exposed to the electron beam, leading to a substantial photoluminescence (PL) background that hinders the identification of B-centers. To address these issues, we pre-cleaned the h-BN flakes in Piranha (see Methods section) prior to electron beam irradiation using a commercial SEM. To avoid surface damage, the irradiation spots were placed close to the edges of the flakes or even at areas outside the flakes, on the Si/SiO$_2$ substrate. This procedure enabled us to create B-centers at random locations up to $20 \mu$m away from the irradiation spot, presumably due to higher-order backscattering of electrons. 

\section{Results}\label{sec2}

\begin{figure}[ht]
    \centering
    \includegraphics[width=0.9\linewidth]{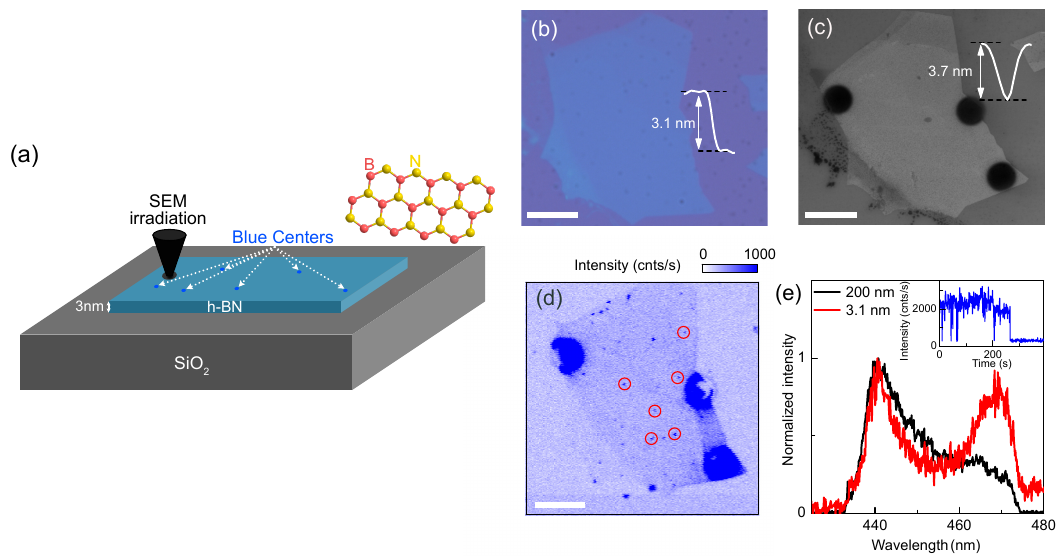}
    \caption{Generation of SPEs in Ultra-Thin hBN: (a) Schematic of the electron beam irradiation setup, where SPEs are also activated far from the irradiation spot. (b) Optical image of the ultra-thin h-BN flake. The inset shows the AFM line cut indicating that the h-BN flake is \qty{3.1}{\nm} thick. (c) SEM image of the flake after 10 minutes of electron beam irradiation. The three dark spots indicate the initial areas of electron scattering from irradiation. The inset of the AFM line cut demonstrates surface damage at the irradiation spots due to the electron beam. (d) PL map measured with an avalanche photodiode (APD), revealing numerous localized SPEs, highlighted by red circles. (e) Room temperature PL spectra, integrated for \qty{30}{\s}, of the SPEs in \qty{200}{\nm} and \qty{3.1}{\nm} thick h-BN. The inset graph shows reduced stability of SPEs in the ultra-thin h-BN. All scale bars represent \qty{10}{\micro\meter}.}
    \label{Figure1}
\end{figure}

\subsection{Remote generation of SPEs in ultra-thin h-BN}

h-BN flakes obtained from high-quality h-BN crystals synthesized by the high-pressure and high-temperature (HPHT) method \cite{taniguchi2007synthesis} are mechanically exfoliated on a \qty{285}{\nm} \(\text{SiO}_2/\text{Si}\) substrate. We then select h-BN flakes with the desired thickness, typically between \SIrange{3}{10}{\nano\meter}, as a platform to create SPEs (Fig. \ref{Figure1} (a)). The thickness is determined from optical contrast and atomic force microscopy (AFM) measurements. Fig. \ref{Figure1} (b) shows the optical image of the typical ultra-thin h-BN flake. An AFM measurement confirms the estimated thickness of \qty{3.1}{\nm} (the white line cut of Fig. \ref{Figure1} (b)). 

Our approach to remotely generate SPEs in ultra-thin h-BN flakes is schematically shown in Fig. \ref{Figure1} (a). To avoid damage to the crystal structure of the h-BN flake (see Supplementary Note 2), we position the electron beam at the edges of the flake. The irradiation voltage and current are consistently kept at 15 kV and 8 nA during the irradiation. Each spot on the flake was irradiated for 10 minutes with a typical electron beam diameter smaller than \qty{10}{\nm}. The SEM image in Fig. \ref{Figure1} (c) shows the dark irradiation spots at the edge of flake where the electron-beam induced damage to the flake, presumably caused by first-order back-scattered electrons. These regions are covered with  $\sim$ \qty{5}{\micro\meter} diameter carbon films that form upon electron beam irradiation.

In previous studies, it was shown that ensembles of emitters or SPEs could be found within the irradiation areas depending on the irradiation dose \cite{fournier2021position}. However, this rather standard method fails for ultra-thin h-BN due to the damage to the surface upon electron irradiation, recently reported in \cite{keneipp2024nanoscale}. The AFM 2D scans of the irradiated flake in Fig. \ref{Figure1} (c) clearly reveal holes etched by the electron beam in the irradiation regions, as indicated by the top white line cut. The estimated depth was \qty{3.7}{\nm} after 10 minutes of irradiation (depth increases with irradiation time, as shown in Fig. S2 b). This depth is even greater than the thickness of the h-BN flake, indicating that the irradiated region has been fully penetrated. This damage is not evident in thick h-BN, but becomes prominent when the h-BN thickness is lower than \qty{10}{\nm}. This is a consequence of the fact that the minimum dose required to create SPEs increases as the h-BN thickness decreases. For \qty{10}{\nm} thick flakes, the required dose is an order of magnitude higher compared to \qty{100}{\nm} thick h-BN flakes.

Following SEM irradiation, we use a home-built scanning confocal microscope (see Fig. S1) to characterize SPEs activated in the flakes. Fig. \ref{Figure1} (d) shows the spatial map of spectrally integrated PL, measured with an avalanche photon diode (APD) following SEM irradiation. The emitters are excited off-resonantly by a continuous wave laser at \qty{405}{\nm} with a fixed power P=\qty{200}{\micro\watt} focused to a diffraction-limited spot. The APD scan shown in Fig. \ref{Figure1} (d) reveals a substantial number of B-centers distributed throughout the flake. We typically select B-centers located more than 10 µm away from the irradiation spots to ensure that they remain free from damage or surface contamination caused by the electron beam. At room temperature, the typical PL spectrum of these emitters in Fig. \ref{Figure1} (e), marked by red circles in Fig. \ref{Figure1} (d), exhibits ZPL at \qty{440}{\nm} and phonon side band at \qty{462}{\nm}.

\begin{figure}[ht]
    \centering
    \includegraphics[width=1\linewidth]{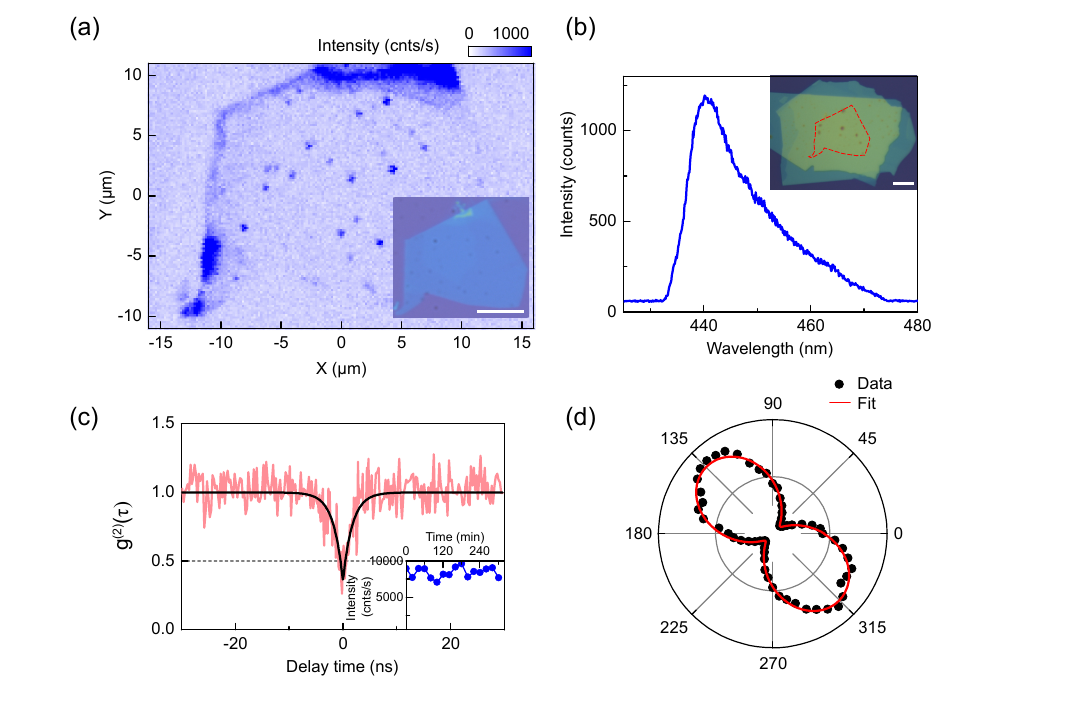}
    \caption{SPEs in an Encapsulated Ultra-Thin h-BN at Room Temperature: (a) PL map of the \qty{6}{\nm} thick h-BN following electron beam irradiation, showing numerous localized SPEs. The inset shows the optical image of the flake, with the scale bar representing \qty{10}{\micro\meter}. (b) PL spectra integrated for \qty{30}{\s} of the SPEs in the irradiated 6 nm h-BN flake and encapsulated in h-BN. The inset shows the optical image of the encapsulated flake, the scale bar represents \qty{10}{\micro\meter}. (c) Photon correlation measurement confirming single photon emission, as indicated by $g^{(2)}(0)<0.5$. The inset graph shows that APD counts of the SPE emission remain stable over 5 hours of constant off-resonant excitation. (d) Polarization analysis of emission from SPEs, indicating a dominant in-plane dipole.}
    \label{Figure2}
\end{figure}

\subsection{Characterization of SPEs in encapsulated, ultra-thin h-BN}

B-centers in ultra-thin h-BN exhibit weaker PL signals compared to those in thicker h-BN, with counts typically about 5 times lower. They become unstable under continuous illumination and are usually photobleached within 30 minutes of exposure to \qty{200}{\micro\watt} laser radiation (see inset of Fig. \ref{Figure1} (e)). We attribute the decreased PL counts, broader spectra, and increased photobleaching of SPE in ultra-thin h-BN to their proximity to the van der Waals surface. To validate this assumption, we encapsulated the irradiated ultra-thin h-BN with \qtyrange[range-units=brackets, range-phrase=--]{30}{40}{\nm} thick layers of h-BN. This approach restored brightness, stability, and narrow linewidth to levels similar to those observed in \qty{100}{\nm} thick h-BN. Fig. \ref{Figure2} (a) shows the PL map of the irradiated 6 nm thick h-BN flake that is later encapsulated, with the optical image of the flake shown in the inset. The map reveals numerous SPEs distributed across the flake, away from the irradiation spots located at the corners. Fig. \ref{Figure2} (b) displays the room temperature PL spectrum of the ultra-thin h-BN flake now encapsulated, with the inset showing its optical image. Fig. \ref{Figure2} (c) shows the photon auto-correlation measurement in a Hanbury-Brown and Twiss configuration of the SPEs, which exhibits anti-bunching behavior with $g^{(2)}$(0) = 0.36 \(\pm {0.04}\). The lifetime extracted from fitting the data with an exponential function is 1.76 $\pm$ \qty{0.17}{\ns}. In particular, the encapsulated B-centers were measured over many hours with no signs of degradation (see the inset of Fig. \ref{Figure2} (c)). Finally, Fig. \ref{Figure2} (d) shows the polarization measurement of the PL, indicating a dominant in-plane dipole with a ratio between the two crossed directions of approximately 10. 

\subsection{Illustration of local dielectric sensing in ultra-thin hBN}

\begin{figure}[ht]
    \centering
    \includegraphics[width=1\linewidth]{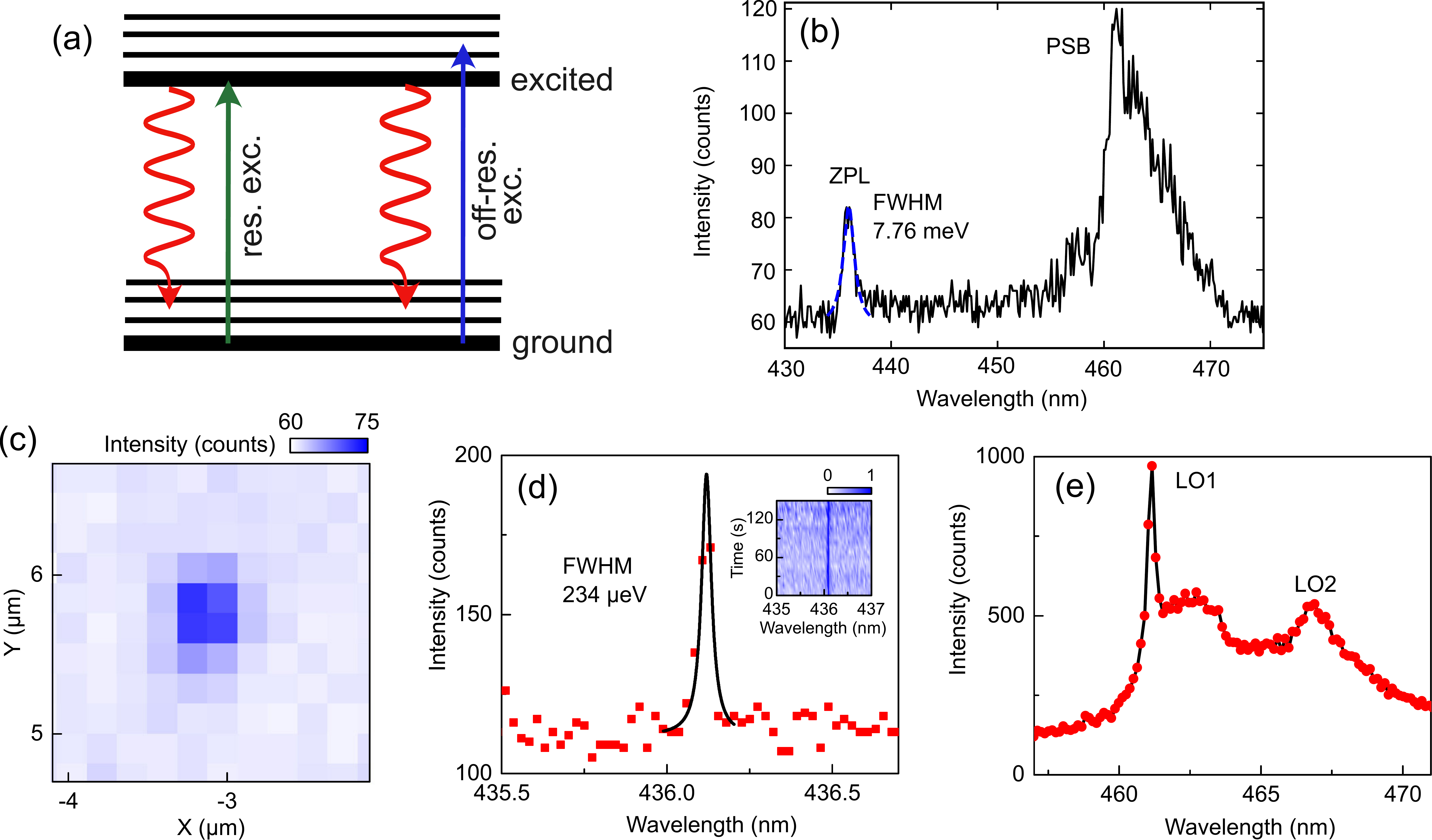}
    \caption{Low Temperature Measurements: (a) Schematic energy level diagram of the B-center, illustrating relaxation pathways under off-resonant (\qty{405}{\nano\meter}) and resonant (\qty{436}{\nano\meter}) excitations. (b) PL spectrum of SPE in the non-encapsulated ultra-thin hBN integrated for \qty{60}{\s} under off-resonant excitation exhibiting a broad ZPL and PSB. (c) Following encapsulation, the integrated PL map within the \qtyrange{434}{437}{\nm} spectral range pinpoints the spatial localization of the SPE. (d) The off-resonant PL spectra integrated for \qty{60}{\s} show ZPL with a width of \qty{234}{\micro\eV}, limited by the spectrometer resolution. The inset confirms no blinking or spectral diffusion of the SPE emission during the observation period. (e) Similar to the behavior in thick h-BN, the phonon sideband (PSB) spectrum, integrated for \qty{30}{\s}, for resonant excitation of encapsulated \qty{6}{\nano\meter} thick h-BN reveals LO1 and LO2 modes at \qty{161}{\milli\eV} and \qty{192}{\milli\eV} above the ZPL, respectively.}
    \label{Figure3}
\end{figure}

Fig. \ref{Figure3} (a) shows a schematic energy level diagram for off-resonant and resonant excitation. Off-resonant excitation at \qty{405}{\nano\meter} promotes carriers to higher electronic states, followed by non-radiative relaxation to the excited state and subsequent radiative emission. Under resonant excitation, our measurements only monitor the optical phonon sideband. We show in Fig. \ref{Figure3} (b) the PL spectrum of SPE in the non-encapsulated ultra-thin hBN measured at \qty{5}{\kelvin} under off-resonant excitation, where both ZPL and PSB were resolved even under off-resonant excitation.  The central wavelength of ZPL emission remains near \qty{436}{\nm}. The measured linewidth of \qty{7.76}{\milli\eV} is substantially broader than of SPEs in thick hBN (see Fig. S6).

Fig. \ref{Figure3} (c) shows the spatial distribution of PL from an SPE in encapsulated h-BN, which is limited by diffraction to $ \sim \lambda$. Fig. \ref{Figure3} (d) displays the PL spectrum at \qty{5}{\kelvin}, where the ZPL is observed at \qty{436.12}{\nm}, corresponding to \qty{2.8429}{\eV}. The extracted linewidth is \qty{234}{\micro\eV}, which is limited by the spectral resolution of the spectrometer. The actual linewidth can be as narrow as several GHz, as measured in photoluminescence excitation measurements \cite{fournier2023investigating,zhigulin2023stark}. No spectral diffusion or blinking is observed over two minutes (see inset of Fig. \ref{Figure3} (d)), indicating that the SPE is immune experiences a clean and stable dielectric environment in the encapsulated ultra-thin h-BN.
These measurements demonstrate that encapsulation of the thin hBN layer hosting the B centers between thicker hBN flakes restores the optical properties of SPEs created in thick hBN, including the narrow linewidth of ZPL. Our technique for generating B-centers through remote e-beam irradiation, therefore, results in negligible surface contamination or damage.

Additionally, we used a \qty{436}{\nm} diode laser to resonantly excite the SPE and collect emission from the optical phonon sideband. Fig. \ref{Figure3} (e) shows two peaks corresponding to longitudinal optical (LO) phonon modes in h-BN \cite{vuong2016phonon}: one at \qty{461}{\nm} with a phonon energy of \qty{161}{\milli\eV}, and the other at \qty{468}{\nm} with a phonon energy of \qty{192}{\milli\eV}. We measured three SPEs, all of which exhibited resolution-limited  ZPL linewidths of less than \qty{300}{\micro\eV} and resonance wavelengths around \qty{436}{\nm} (see Fig. S4). Furthermore, we observed stable SPEs in a second encapsulated ultra-thin h-BN sample (see Fig. S5).

\section{Discussion}\label{Discussion}

The principal goal of our work was to develop SPEs within h-BN flakes as thin as \qty{3}{\nm}, thus enabling precise control over the vertical positioning of defects. By harnessing the atomic-scale nature of these quantum emitters, we envision integrating ultra-thin h-BN SPEs into heterostructures in close proximity to targeted quantum materials. This approach aims to enable imaging of quantum phases emerging in 2D systems with sub-nanometer spatial resolution, by exploiting the sensitivity of the ZPL resonance to electric fields through the DC-Stark effect \cite{zhigulin2023stark}.

In summary, we have successfully created SPEs in ultra-thin h-BN. The non-invasive method we developed for creating SPEs avoids contamination and surface damage. It offers a high yield, is easy to operate, does not require temperature annealing of h-BN flakes, and can be completed within a single day at a standard research facility. The proposed method for fabricating SPE in ultra-thin h-BN, however, inherently lacks the precision for controlling their lateral spatial positioning, as the emitters are created outside the irradiation spots.  We confirm single photon emission at room temperature through observation of photon anti-bunching, with $g^{(2)}(0)$ = 0.36 \(\pm {0.04}\) and a resolution-limited linewidth measured at 5 K of less than \qty{300}{\micro\eV}.

Our findings reveal that the brightness, stability, and spectral characteristics of SPEs in ultra-thin h-BN are adversely affected by an unstable dielectric environment.  When these SPEs are encapsulated in thicker h-BN layers, their performance is restored to levels comparable with those observed in thicker h-BN. This restoration highlights the extreme sensitivity of SPEs in ultra-thin h-BN to their dielectric surroundings, emphasizing their potential as highly responsive sensors that could ensure unprecedented spatial resolution in detection of electric fields. In particular, we envision that B-centers in ultra-thin hBN transferred onto tan AFM tip could be used to obtain sub-nanometer resolution imaging  of Wigner crystals through periodic electrostatic potentials they generate.

\section{Methods}\label{Methods}
\subsection{Fabrication}
The HPHT h-BN crystal is sourced from the National Institute for Materials Science (NIMS). Thin layers of h-BN were obtained via mechanical exfoliation on SiO\(_2\) (\qty{285}{\nano\meter})/Si. The thickness of h-BN was measured using optical reflection contrast and atomic force microscopy. To remove polymer residuals and hydrocarbon deposits introduced during exfoliation and from the ambient environment, the samples were treated with Piranha solution for 20 minutes at room temperature. The Piranha solution is composed of \qty{10}{\milli\liter} \( \text{H}_2\text{O}_2 \) (\qty{30}{\%}) and \qty{40}{\milli\liter} \( \text{H}_2\text{SO}_4 \) (\qty{95}{\%}). The clean samples were then immediately irradiated with electron beams using a Zeiss ULTRA 55 plus. The acceleration voltage is \qty{15}{\kilo\volt} and the current is \qty{8}{\nano\ampere}. The size of the irradiation spot is less than \qty{10}{\nm}. Our method eliminates the need for temperature annealing of h-BN flakes.

\subsection{Optical Measurements}
The irradiated samples were characterized using a home-built confocal microscope. The excitation wavelengths of either a \qty{405}{\nm} or \qty{436}{\nm} continuous-wave diode laser were used for non-resonant or resonant photoluminescence (PL) measurements. The laser beam was focused on the samples with a 0.85 NA objective at room temperature or a 0.55 NA objective at low temperature. The emitted PL signals were filtered with a bandpass filter ($450\pm20$ nm) and collected by avalanche photodiode single-photon detectors (Excelitas Technologies) or a spectrometer (Princeton Instruments, Inc.) of which the highest resolution achieved using \qty{1500}{g~mm^{-1}} optical grating. A galvanic mirror pair from Thorlabs, Inc. in combination with a 4f optical system enabled precise two-dimensional scanning of the samples. A bath cryostat and a helium flow cryostat (Oxford Instruments) with a base temperature of \qty{4.2}{\kelvin} was used for low-temperature measurements. 
The bin size of the time-correlated single photon counting (PicoHarp 300) was \qty{128}{\pico\second}. The integration time for the spectral measurements was set to 30 s, except for Fig. \ref{Figure3} (c), where it was extended to 60 s. The spectrum shown in Fig. \ref{Figure3} (b) was measured using a lower-resolution \qty{300}{g~mm^{-1}} grating to capture the broader spectral features.

\backmatter

\bmhead{Supplementary information}
The online version contains supplementary material available.

\bmhead{Acknowledgements}
We are grateful to Aymeric Delteil for sharing his know-how on h-BN defects. We also thank Shengyu Shan, Jiecheng Feng and Eugene Demler for stimulating discussions.
This work was supported by the Swiss National Science Foun-
dation (SNSF) under Grants No. CRSII-222792 and 200021\_204076.

IK acknowledges the financial assistance of the Rothschild Post-Doctoral Fellowship from Yad HaNadiv, the Helen Diller and the Viterbi Post-Doctoral Fellowships from Technion.

K.W. and T.T. acknowledge support from the JSPS KAKENHI (Grant Numbers 21H05233 and 23H02052), the CREST (JPMJCR24A5), JST and World Premier International Research Center Initiative (WPI), MEXT, Japan.

\bmhead{Author contributions}
AI, MK, IK and LL conceived the project. LL, IK, JE, BR and SF fabricated the samples. LL, IK and JE, with assistance from MK, conducted the experiments and analyzed the data. All authors contributed to discussions and manuscript preparations.

\section*{Declarations}
The authors declare no competing interests.

\bibliography{sn-bibliography}

\end{document}